# Review of Core Process Representation in Power System Operational Models: Gaps, Challenges, and Opportunities for Multisector Dynamics Research

Konstantinos Oikonomou[1], Brian Tarroja[2,3], Jordan Kern[4], Nathalie Voisin[1,5]*

[1]*Pacific Northwest National Laboratory, Richland WA*
[2]*Department of Civil and Environmental Engineering, University of California Irvine*
[3]*Advanced Power and Energy Program, University of California Irvine*
[4]*Department of Forestry and Environmental Resources, North Carolina State University*
[5]*Civil and Environmental Engineering, University of Washington, Seattle, WA*

*corresponding author: Nathalie.voisin@pnnl.gov

## Abstract

Power grid operations increasingly interact with environmental systems and human systems such as transportation, agriculture, the economy, and financial markets. Our objective is to discuss the modelling gaps and opportunities to advance the science for multisector adaptation and tradeoffs. We focus on power system operational models, which typically represent key physical and economic aspects of grid operations over days to a year and assume a fixed power grid infrastructure. Due to computational burden, models are typically customized to reflect regional resource opportunities, data availability, and applications of interest. While there are model intercomparison papers, there is however no model-agnostic characterization and systematic overview of the state-of-the-art process representations in operational power system models. To address our objective, we conceptualize power system operational models with four core processes: physical grid assets (generation, transmission, loads, and storage), model objectives and purpose, institutions and decision agents, and performance metrics. We taxonomize the representations of these core processes based on a review of 23 existing open-source and commercial models. As we acknowledge the computational burden of certain representations, we leverage this taxonomy to describe tradeoffs in process fidelity and tractability that have been adopted by the research community to address interactions between the power grid and hydrometeorological uncertainties, global change, and/or technological innovation. The core process taxonomy along with the existing computational tradeoffs are used to identify technical gaps and recommend future model development needs and research directions to better represent power grid operations as part of integrated multisector dynamics modeling and interdisciplinary research.

## Highlights

- A novel process-based taxonomy for power system operational models is developed

- Taxonomy of core process representations is based on the review of 23 models

- We discuss key core process representation and computational tradeoffs by science question

- We identify technical gaps and opportunities to support multisector dynamics research

## 1. Introduction

The electricity sector is undergoing a technological and institutional transformation: deep penetration of variable renewable energy (VRE); widespread adoption of energy storage (ES) technologies and electric vehicles (EVs); potential future regulatory and institutional changes like a carbon tax; increased participation in demand-side management programs; and re-design and geographical alignment of electricity markets (1). This transformation is influenced by other sectors (2), such as water (3-5), which has been undergoing its own transformation and water crisis for decades due to growing demands and increasing competition between water uses. (6) developed the socio-hydrology concept to characterize this crisis and direct research in transformation, while (7) defined water as a master variable for the transformation of other sectors of activities (e.g., agriculture and energy). In view of those dependencies, recent studies have explored the interactions with other sectors and water in particular in more quantitative ways (8-12). Understanding how changes in water systems, technologies, transportation, communication, governance, and markets will affect power systems in the future, and how the impacts could cascade across sectors, represents an outstanding research challenge. To meet this challenge, the research and modeling community must develop best practices for linking new and/or existing operational models of power systems within larger, integrated, multisector modeling frameworks such as those developed by (Kraucunas et al. 2015) and (13). However, relatively little formal guidance exists for operational model selection and/or development within the specific context of multisector modeling.

Power system models can be divided into three general classes depending on their scope, spatial resolution, and end use application (Figure 1):

- Dynamic models, i.e., "power flow" or "stability" models;
- Planning models, i.e., "capacity expansion" models;
- Operational models, i.e., "production cost" or "unit commitment/economic dispatch" (UC/ED) models.

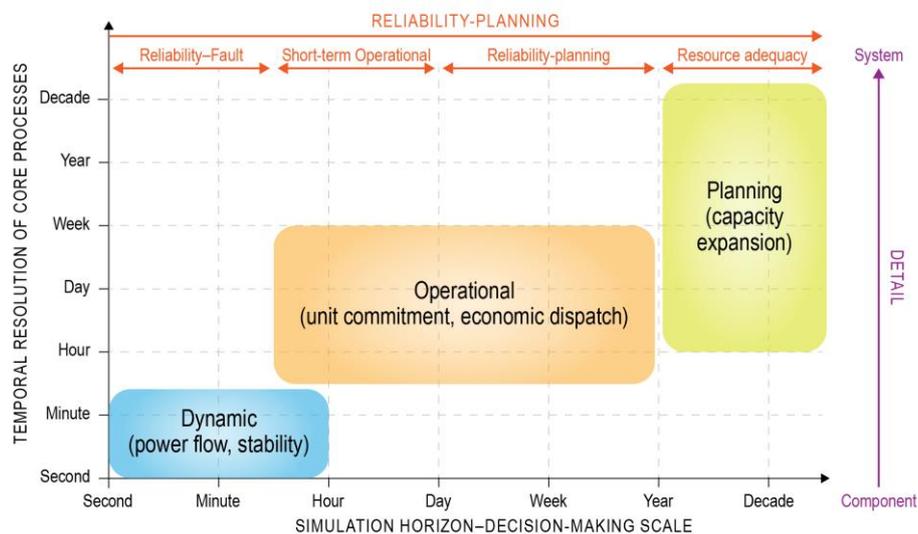

Figure 1: Illustration of application versus temporal resolution and temporal horizon.

Dynamic models perform detailed static load flow (steady state) and dynamic power flow simulations over short time periods (~ 30 sec to 1 min), and are used primarily to examine cascading outage events and/or the ability of existing systems to recover (e.g., generator tripping) (14).

Long-term planning models optimize capital investment in power system infrastructure and are typically used in global integrated assessments and integrated resource planning, that power utility organizations are required to perform routinely (11, 13, 15-20). These models are not intended to simulate time-resolved system behavior over long continuous periods and typically rely on snapshots representing different conditions in the simulation period.

Operational models are used to provide the spatiotemporal specificity of how a bulk power system might be operated. These models act on time scales that bridge the temporal resolution of dynamic models and planning models. Similar to dynamic models, operational models assume a fixed topology of generation and transmission assets, and simulate the performance of the modeled system using locational time series of hourly or sub-hourly electricity demand and VRE production. Compared to planning models, operational models provide higher-resolution understanding of system dynamics and overall performance (including market prices and vulnerabilities to extreme events).

Research evaluating long-term planning models and how the representation of other sectors influences the energy sector is emerging (21-23), but little guidance exists about modeling choices for fidelity[a] and tractability[b] in power system operational models to support advances in multisector research (24, 25). For example, (26) reviewed over 75 operational power system models in their ability to evaluate the integration of renewables; (27) reviewed representations of vehicle electrification; (28) reviewed seven proprietary models and their responses to market structures. Yet, there is no model-agnostic characterization and systematic overview of the process representations in operational power system modeling to identify multisectoral modeling gaps in power system models and understand how the potential vulnerabilities to other interdependent sectors (water, transportation, agriculture, markets, etc.) may influence the operations of the grid.

---

[a] Fidelity is the degree to which an operational model reproduces the state and behavior of a real-world power system (e.g., time resolution, network detail, geographic scope, characterization of uncertainty) in a measurable or perceivable manner.
[b] Tractability refers to the computational complexity and time required for an operational model to solve an operational problem.

In order to address modeling gaps in a model-agnostic way, we develop a novel process-based taxonomy of power system operational models and provide guidance about the underlying model structures and process representations needed to better enable the investigation of multisector science questions. We aim to guide modeling choices on par with the science questions, availability of data, computational needs, and overall applications. Our intention is for this paper to be used as a general blueprint for future development of both stand-alone and integrated open-source operational models and overall guidance for documentation and evaluation by the multisector modeling and research community.

The rest of this paper is organized as illustrated in Figure 1 and as follows: In Section 2, we develop a taxonomy of core process representations from 23 existing and representative open-source and commercial operational models. We acknowledge that each reviewed model was customized for specific regions and science questions with a keen view on computational burden. In Section 3, we introduce several examples of multisector science questions involving simulation of power system operations and report on computational tradeoffs in the choice of core process representations. Leveraging the taxonomy of core process representations and modeling choice tradeoffs, in Section 4 we identify modeling gaps that are key to integrate multisector concepts. We discuss opportunities and approaches necessary for overcoming some of these gaps and deficiencies in order to support multi sector research.

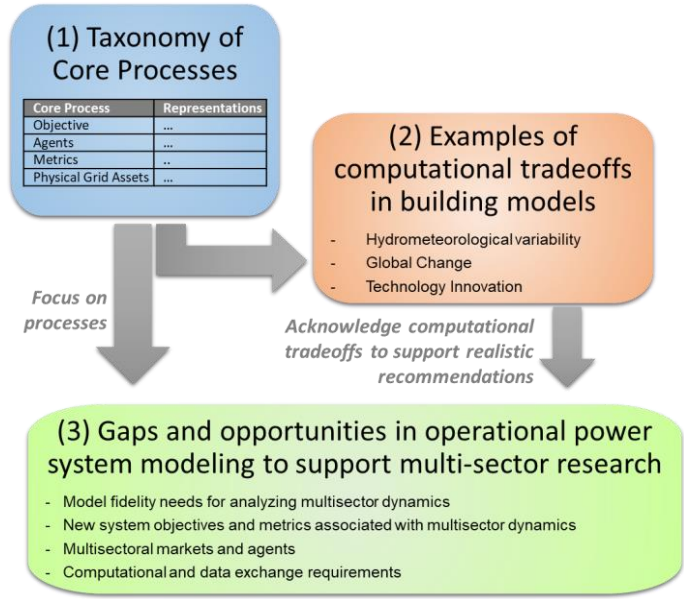

Figure 1: Review approach to support the discussion on future research directions and development needs.

## 2. Taxonomy of Existing Modeling Representations

We conceptualize power system operational models with four core processes that are critical for capturing interactions between the power system and other sectors: (1) physical grid assets (transmission, generation resources, loads); (2) model objectives and purpose which coordinate and manage the operations of the physical grid assets according to certain rules and policies; (3) institutions and decision agents which decide on the rules and policies; and (4) performance metrics which measure the performance of the overall coordination of individual physical assets in following rules and policies (Figure 2). We apply the taxonomy to 23 existing operational models (see Table S1). We will discuss implications for computational tractability and model fidelity to real systems in Section 3 for simplicity and clarity. Table 1 below gives an overview of the taxonomy and the total number of the reviewed models associated with each modeling representation. Each element of the table is briefly analyzed below starting with the model objective function.

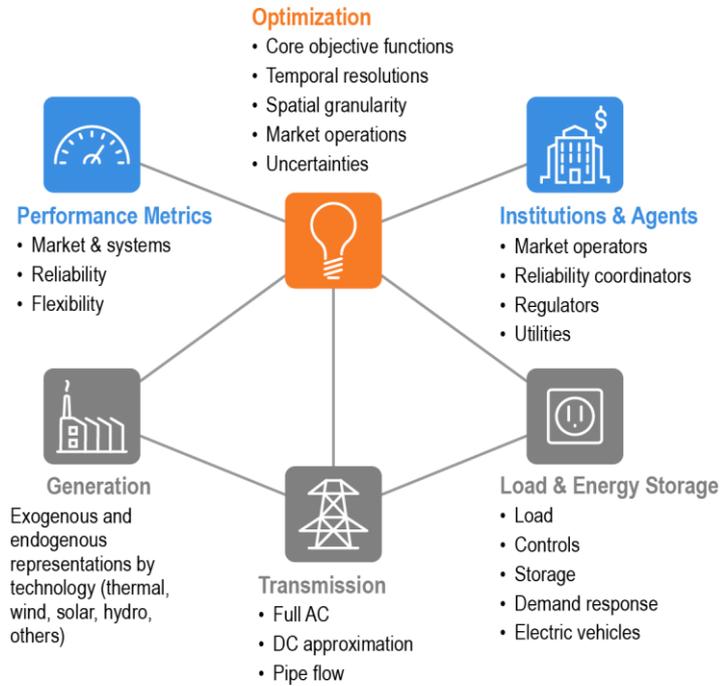

Figure 2: We conceptualize operational models in 4 core processes: (1) physical grid assets in grey (i.e., generation resources, transmission, load and energy storage resources); (2) optimization that coordinate the operations of the physical assets according to (3); (3) institution and agents; and (4) performance metrics.

Table 1 below gives an overview of the taxonomy and the total number of the reviewed models associated with each modeling representation. Each element of the table is briefly analyzed below starting with the model objective function.

Table 1: Taxonomy of core processes

| Core Process | Modeling Representations | Modeling Sub-Representations | Number of Models (out of 23) |
|---|---|---|---|
| Physical Grid Assets: Spatial Granularity | Zonal | | 23 |
| | Nodal | | 11[c] |

---

[c] By default, all zonal operational models can also perform nodal simulations

| Physical Grid Assets: Transmission Line Constraints | AC[d]- Urban Scale | | 3[e] |
|---|---|---|---|
| | DC[f] - BPS[g] Scale | DC Approximation | 12 |
| | | Pipe Flow | 10 |
| Physical Grid Assets: Generation Resources | Thermal | Baseload | 23 |
| | | Intermediate | 23 |
| | | Peaking | 23 |
| | Hydro | Water model exogenous unit commitment | 6 |
| | | Unit commitment based on monthly hydropower potential – Default (requires HTC[h] or PLF[i] parameterization – Dispatch optimized by the power system model, limited water constraints) | 9 |
| | | Hybrid water model and power unit commitment model – Intermediate (weekly to daily energy potential requires HTC and/or PLF – dispatch optimized by the power system model, constrained by water availability at relevant time scales) | 3 |
| | | Fixed water schedule | 5 |
| | Variable Renewable Energy Resources | Must-Take | 19 |
| | Distributed Generation resources | Must-Take | 13 |
| Physical Grid Assets: Load and Energy Storage Resources | Plugged-in Electric Vehicles | Must Serve or deferrable loads | 11 |
| | Interruptible Loads | Can be curtailed as needed | 11 |
| | Energy storage: pumped hydro storage, compressed air energy storage, concentrating solar power storage, electrochemical batteries, flywheels | Given Schedule | 8 |
| | | Peak Shaving | 9 |
| | | Price Driven | 19 |
| | Objective Function | Maximize Profits | 3 |

[d] Alternating Current
[e] By default, all zonal operational models can also perform nodal simulations
[f] Direct current
[g] Bulk power system
[h] Hydro-Thermal Coordination
[i] Proportional Load Following
88

| | | | |
|---|---|---|---|
| Models Objective and Purpose | | Minimize Operating Cost | 18 |
| | | Other | 2 |
| | Temporal Resolution | Hourly | 23 |
| | | Sub-hourly | 11[j] |
| | Optimization Horizon | Day, week, year, multiple years | 23 |
| | Uncertainties | Stochastic | 8[k] |
| | | Deterministic | 23 |
| Institutions and Decision Agents | Market operators (including all market players: electric utilities, load service entities, electricity traders) | Day-Ahead (DA) market | 23 |
| | | Real-Time (RT) market | 9[l] |
| | Reliability coordinators (e.g., WECC, generation and transmission planners) | | 14 |
| | Regulators (e.g., state energy offices, public utilities commissions, FERC) | | 2 |
| Performance Metrics | Market and System Metrics | Hourly dispatch of generating resources | 23 |
| | | Production cost of the bulk power system | 21 |
| | | Locational marginal prices | 17 |
| | | Power flows on transmission lines, transmission losses, congestion cost of limiting lines, and voltage angles. | 21 |
| | | Electricity producers' payments and costs | 21 |
| | Reliability Metrics | Emissions and emission costs | 13 |
| | | Wind/solar/hydro spillage | 21 |
| | | Energy not served | 7 |
| | | Energy not served cost | 7 |
| | | Loss of Load Expectation | 6 |
| | | Loss of Load Probability | 7 |

---

[j] By default, all ten operational models that perform sub-hourly simulations can also perform hourly simulations
[k] By default, all seven stochastic operational models can also perform deterministic simulations
[l] By default, RT operational models can also perform DA simulations

|  | Flexibility Metrics | Reserve inadequacy | 1 |
|  |  | Unused hydro reservoir capacity | 1 |
|  |  | Residual ramping capability | 1 |

## 2.1 Model objectives and purpose

Power system operational models often belong to a family of mathematical optimization problems, and most of their core decision variables control the commitment status (i.e., on/off) and dispatch (i.e., amount of generation) of generators. These are subject to unit-specific constraints (e.g., maximum capacity, minimum up and down times, ramping limits) and system-wide constraints (e.g., meeting demand for electricity and ancillary services, transmission line limits, etc.). Common objective functions include (1) minimize total power system operating costs (least-production cost); (2) maximize profits from energy production; (3) maximize the uptake of renewable or zero-carbon energy (and other, non-monetary objectives), as explored in more detail below.

### 2.1.1 System-wide cost minimization

This representation minimizes the aggregate, system-wide cost of meeting electricity demand and ancillary service requirements. These costs largely consist of fuel costs and start-up/shut-down costs of thermal generating units. Ancillary services refer to reliability measures (e.g., unused capacity) provisioned by generators and flexible load resources that enable grid operators to take corrective actions in real-time operations such as sudden generating unit or interconnection outages (29). Emission constraints (e.g., the total amount of a pollutant such as $CO_2$, $SO_2$, $NO_x$ released over a year) can be added as part of the optimization process to improve air quality standards (30).

### 2.1.2 Firm or asset profit maximization

In the profit maximization representation, independent power producers (IPPs) seek to optimally schedule the timing and level of power production from generators based on exogenously defined time series of energy and ancillary service prices. This representation generally tends to focus more on the operational objectives and constraints of individual generating units, and assumes that IPPs are "price takers" without having control over electricity market prices (31).

### 2.1.3 Non-monetary objectives

Operational models can also be formulated around objectives that are not explicitly financial or economic. These types of models are often used for more exploratory research rather than real-time decision support. They can offer added insight into how system performance changes when alternative objectives (e.g., maximizing the uptake of renewable or zero-carbon energy (32) or minimizing greenhouse gas [GHG] or criteria pollutant emissions) are used to drive the optimization. Typically, these models adhere to the technical constraints of generators, transmission systems, and flexible loads and satisfy the criterion of adequate supply for meeting the demand for energy and ancillary services.

### 2.1.4 Features of optimization

#### 2.1.4.1 Optimization time step and horizon

Operational models generally employ an hourly time step and optimization horizons that span a single day (24 hours), week (168 hours), or year (8760 hours). To better capture the variability in systems with a large share of renewable energy resources where ramping needs might exceed the capability of dispatchable thermal units, some operational models also offer intra-hourly simulations, ranging from five minutes to half an hour (33), albeit typically using shorter scheduling horizons (e.g., a day).

*2.1.4.2   Stochastic vs. deterministic optimization*

Operational models generally take one of two approaches for incorporating uncertainty into the optimization procedure: deterministic or stochastic. In deterministic operational models, grid participants are assumed to have perfect foresight with respect to changing values of electricity demand, VRE production, unexpected system failures, and electricity prices over a given horizon. Uncertainty in these processes and their associated effects can be explored using Monte Carlo analysis, but it is not explicitly part of the underlying optimization (34).

In stochastic operational models, a two-stage optimization approach is usually considered (35). First, the model makes initial commitment and dispatch decisions for all generating units based on generators' initial conditions and an ensemble of possible forward-looking scenarios. Second, the model advances a single time step, and some portion of future operating conditions is resolved. The process is then repeated, and the operational model finds the optimal adjustments to the initial schedule based on realized forecast errors and updated information for future days. In the case of operational models with explicit representation of hydroelectric dams, stochastic dynamic programming (SDP) or stochastic dual dynamic programming (SDDP) methods can be used to integrate uncertain parameters (e.g., streamflow) into the optimization process (36), as discussed in Section 3.

## 2.2   Institutions and decision agents

Decision agents add contextual nuance to the design and use of power system operational models and influence the model purpose and metrics. Relevant grid participants that may need to be explicitly represented in operational models include system operators, power utilities, and IPPs.

### 2.2.1   Grid operators

Grid operators, or regional transmission organizations (37) and independent system operators (ISOs), are nonprofit operators that ensure the reliable operation of regional electricity grids and administer associated wholesale electricity markets. Grid operators are tasked with organizing available hourly generating resources (typically on a dollar per megawatt cost basis) to solve a cost minimization problem that "clears" wholesale electricity markets (schedules specific plants to run and determines a market price). ISOs/RTOs typically oversee the execution of "day-ahead" and "real-time" decision-making processes:

- Day-ahead energy and ancillary service markets open several days before the commitment period (e.g., a week) and close a day ahead. Day-ahead markets allow participants to buy or sell wholesale electricity before the operating day to minimize reliability issues and price volatility.
- Real-time energy and ancillary markets operate on much shorter intervals, clearing every 15 or 5 minutes, and act as a balancing market where the day-ahead commitments are balanced against actual demand and system constraints, accounting in many cases for forecast errors.

Operational models are typically structured to represent the viewpoint of ISOs/RTOs through their focus on optimizing the centralized dispatch of generators and modeling the electricity markets and their participants (38).

### 2.2.2 *Power utilities and independent power producers*

In jurisdictions without competitive wholesale electricity markets, operational models are used to represent the operations of vertically integrated power utilities. Like ISOs/RTOs, electric utilities typically solve least-production cost optimization problems to determine the optimal daily and hourly commitment and dispatch schedules for their portfolio of generating units. In systems with competitive electricity markets, power utilities are often represented as electricity market participants that can buy and sell power, depending on their resources and internal electricity demands, while IPPs are typically represented as agents that exclusively sell power and ancillary services to maximize profits.

## 2.3 Performance metrics

Performance metrics are the key outputs of operational power system models and can be categorized as (1) market and system; (2) reliability; (3) and flexibility metrics (Table 1). They reflect physical assets' individual and collective contributions to grid operations, exogenous forcings acting on the grid, and the relevant institutions involved. In terms of multisector modeling, these metrics represent the most direct way to measure the potential for power system dynamics to spill-over and affect outcomes in other sectors, and vice versa. The ability of the model to represent metrics associated with other sectors also reflects its multisector fidelity and relevance for co-management.

## 2.4 Representation of physical grid assets

Physical grid assets represent the actual bulk power system managed by the optimization under guidance from the agents. They include generating units, electricity demand locations and the lines to transport electrons to the demand centers. The topology of the bulk system as well as the operational fidelity are described below while tradeoffs are discussed in section 3.

### 2.4.1 Spatial granularity (zonal vs nodal)

Operational models can represent electricity networks with different levels of granularity, typically either nodal or zonal. A nodal representation tries to mimic the physical structure of the electricity network, including the location of all substations and transmission lines that connect them. This level of modeling detail requires substantially more input data (e.g., loads connected at individual nodes, transmission line characteristics), which may be challenging to collect and involve longer computation times. Zonal representation simplifies electricity networks by aggregating multiple demand nodes into larger zones and, similarly, grouping transmission lines between zones into equivalent inter-zonal links. Intra-zonal lines (and transmission congestion) are often omitted from consideration (39).

## 2.4.2 Transmission line representation (AC, DC, and "pipe flow")

The flow of power through transmission lines is influenced by each line's characteristics (i.e., capacity) and the physics of electricity flow (40). The three commonly used power flow models are the alternating current (AC) power flow, direct current (DC) power flow, and "pipe flow" models. The power flow constraints of the DC power flow model are a linearized version of the constraints of the non-linear AC power flow model. The DC approach is less computationally intensive and is widely used by many commercial operational models that simulate the economic operation of large-scale bulk power systems. However, it only considers active power flows and assumes perfect voltage support and reactive power management. An even simpler approach is the pipe flow model, which limits power flows only by transmission line capacity and completely neglects network losses.

The impact of transmission line constraints is particularly important in operational models where the accurate modeling of power lines' congestion and losses is essential to capture system outcomes (e.g., price dynamics) (41).

## 2.4.3 Generation resources

Operational models consider a wide variety of generation resources that can be broadly categorized as thermal (nuclear, coal, natural gas, geothermal), hydroelectric, and VRE (wind, solar) generating units.

### 2.4.3.1 Thermal units

Thermal generating units convert thermal energy released from the combustion of fossil fuels (e.g., coal or natural gas), from nuclear fission reactions, or geothermal heat input to electricity using steam or combustion turbines (42). The efficiency with which the unit converts chemical energy embedded in fuel to electricity is measured by the unit's "heat rate" (MMbtu/MWh), which represents the fuel input needed to produce a given unit (usually 1 kWh or MWh) of electricity output (43). Thermal generating units with lower heat rates have greater efficiency, lower marginal costs, and typically are dispatched first (i.e., more frequently) in least-cost optimization problems.

Apart from nuclear and geothermal power plants, thermal generators are typically treated as "dispatchable" resources, meaning that their power output can be adjusted or turned on/off at the request of ISOs/RTOs or IPPs, according to market and/or reliability needs. Some operational models further categorize thermal generating units as baseload, intermediate, and peaking units depending on their technical characteristics (i.e., ramp rate and minimum up/down times), which vary based on turbine technology, fuel type, and capacity factor (38).

### 2.4.3.2 Hydropower

Hydroelectric generating units convert hydraulic power (gravitational potential energy of flowing water) into electrical energy. Hydropower production is limited by water availability and multi-objective water management priorities such as flood control, water supply, environmental flows, navigation, and recreation. We synthetize the wide range of representations into four categories:

1. Water model exogeneous unit commitment: In this representation, a river-routing reservoir operations model optimizes individual power plant operations according to water management objectives and in response to projected electricity prices, or dynamic prices in the case of co-optimization models. The hourly generation at individual hydropower plants is then input to the power system operational model and is considered as "must-take" by the unit commitment process at each time step. The river-routing reservoir operations model and power system operational model can be run in sequence with a prescribed frequency (hourly or daily) to update information (price and generation) over the operating horizon and not overly constrain the power system model dispatch (44, 45). Co-optimization models look for a dual optimum on both water and electricity objective functions within the same time step.

2. Unit commitment based on monthly hydropower potential and limited water constraints (referred as default approach thereafter): In this representation, hydropower potential is exogenously specified and endogenously scheduled by operational models in one of two ways:
   a. Proportional load following (PLF): A portion of the hydropower generation available is assigned to follow the system's load shape within provided minimum and maximum generation capacity and energy limits. A proportionality constant (usually referred to as "K" value) is calibrated at each plant, reflecting the plant's ability to adjust to load variations.
   b. Hydrothermal coordination (HTC): A portion of individual hydropower plant's available generation is dispatched in accordance with the overall model's cost minimization objective (meaning hydropower is scheduled during the highest marginal cost hours). The HTC representation relies upon the PLF generation schedule, which is further modified by a proportionality constant (usually called the "p" factor). This factor represents the fraction (between 0 and 1) of a plant's capacity that can adjust its output based on market price (46).

   Both PLF and HTC representations rely on hydropower availability first being defined outside the core power system optimization process. In large spatial domains with limited data availability, hydropower potential might be provided at individual or a small group of power plants at a monthly time scale, along with specified hourly ramping rates, and minimum/maximum hourly generation values.

3. Hybrid water model and power unit commitment model (referred as intermediate approach thereafter): This representation is intermediate between the default approach and the water model exogeneous unit commitment approach. In this representation, typically offline water models provide daily and weekly hydropower potential and corresponding hourly operational constraints associated with the water conditions during that period. The travel time between power plants can be added to further constrain the commitment process. The hydropower scheduled is optimized by the operational model according to the system needs with the PLF and HTC representations.

4. Fixed water schedule: For power system operational models with no representation of hydropower unit commitment, a time series of either hourly hydropower generation schedules, or daily schedules with hourly profiles, is provided to the model. The hydropower dispatch is typically either derived from observation or exogenously optimized by water models and considered to be must-take. This representation is static and does not respond to changes in power system dynamics such as prices and reserve needs at any time during the power system simulation. This approach also complements the two latter representations with the hydropower schedule optimized by the power system core optimization process, because some plants—e.g., run-of-the-river and sometimes cascading systems—are too constrained in reality to be represented by default and intermediate representations.

### 2.4.3.3 Wind and solar

Solar and wind generating units are the most common VRE resources found in operational models. Their generation output is intermittent, meaning they have a variable profile controlled by meteorological conditions (e.g., solar irradiance and wind speed). As a result, operational models treat solar and wind generators as non-dispatchable sources of zero variable costs, but both can be curtailed if necessary. Solar and wind generation data (either deterministic or forecasted) are imported as hourly inputs into operational models.

In recent years, there has been a notable increase in generation capacity from distributed generation resources (DGs) connected to power distribution networks (47). DGs include behind-the-meter renewable generators (small-scale solar or wind generators that individual customers install to avoid purchasing electricity from a power utility). DGs are typically modeled similarly to renewable energy resources by most operational models; the hourly generation values are aggregated on a regional level and are subtracted from the regional load.

*2.4.4 Energy storage*

Operational models can be used to model a range of energy storage (ES) devices, including but not limited to (1) pumped hydroelectric storage, (2) compressed air energy storage, (3) concentrating solar power storage, and (4) electrochemical batteries. ES technologies are represented as separate generating and charging units that interact with a state of charge intertemporal constraint including electric losses and roundtrip efficiency (48, 49). The management of this stored potential can have several distinct representations in operational models (50):

1. Fixed schedule: Time series of charge (load) and discharge (generation) profiles are pre-specified by the user and inputted to operational models.
2. Peak-shaving: Charge and discharge schedules are optimally allocated via the core optimization processes to flatten (smooth out) a given daily or weekly load profile subject to storage constraints.
3. Price-driven: Charge and discharge schedules are optimally allocated to respond to market prices subject to the storage technology specification constraints.

*2.4.5 Demand for electricity and ancillary services*

*2.4.5.1 Electricity demand*

Electricity demand has traditionally been treated as a time-varying (usually hourly or sub-hourly) exogenous input. Grid participants use a variety of methods to forecast short-, medium- and long-term electricity demand patterns, including regression and time-series models, neural networks, statistical learning algorithms, fuzzy logic, and expert systems (51).

Increasingly, however, electricity demand is becoming more flexible (i.e., able to be reduced temporarily to address physical and/or economic contingencies)—a resource referred to as demand response or flexible loads (52, 53) Capturing operator control or incentivization to harness this flexibility is important in operational models. Examples include but are not limited to:

1. Building demand response: where appliances and air-handling equipment in buildings are controlled within the constraints of meeting their main function (i.e., maintaining internal space temperatures within a given threshold). This may entail full load shaping or simple curtailment of loads when costs or emissions are high (54).
2. EV charging: where the load associated with charging EVs is shaped by grid operators, within the constraints of local charging infrastructure power limits and the need to provide enough net charge for their drivers to meet their travel needs (55) In some cases, EVs can also discharge electricity back to the grid and act as a form of battery energy storage.
3. Fuel production loads: Equipment such as electrolyzers that produce fuel such as hydrogen for industrial applications can be operated to maximize fuel production during low-cost or low-emissions time periods, within the constraints of on-site storage and meeting the bulk demand for that fuel (56).

### 2.4.5.2 Reserve requirements

Reserve requirements are commonly treated as exogenous inputs based on the "N-1" security criterion. To comply with the N-1 criterion, adequate generation capacity in the form of ancillary services needs to be available throughout the system to prevent loss of load when the single largest generation unit or transmission line is unexpectedly disconnected from the power grid (57). In practice, reserve requirements are computed as a percentage of the total forecasted electricity demand aggregated on a regional (balancing authority [BA]) or power system level (58).

The novel taxonomy of individual core process representations presented above is used in the next section to discuss their combinations within particular operational models and associated modeling choice tradeoffs.

## 3. Example of Computational Tradeoffs

In this section, we use three typical power system science questions to illustrate the modeling tradeoffs of core process representations: (1) operating under hydrometeorological variability; (2) exploring opportunities and vulnerabilities under global change; and (3) evaluating the adoption of new technologies (Figure 3). The three examples are representative, not exhaustive. They also have a multisector component for the specific purpose of our review. Please note that this section neither intends to choose an individual operational model nor to synthesize the optimal combination mix of core process representations to best answer a particular science question. Instead, we report the most common combinations and related modeling tradeoffs of core process representations when operational models are utilized to study the science questions. This section informs on computational burden realities to provide a pragmatic support when discussing modeling gaps and enhancement opportunities to advance our understanding of co-evolving systems in Section 4.

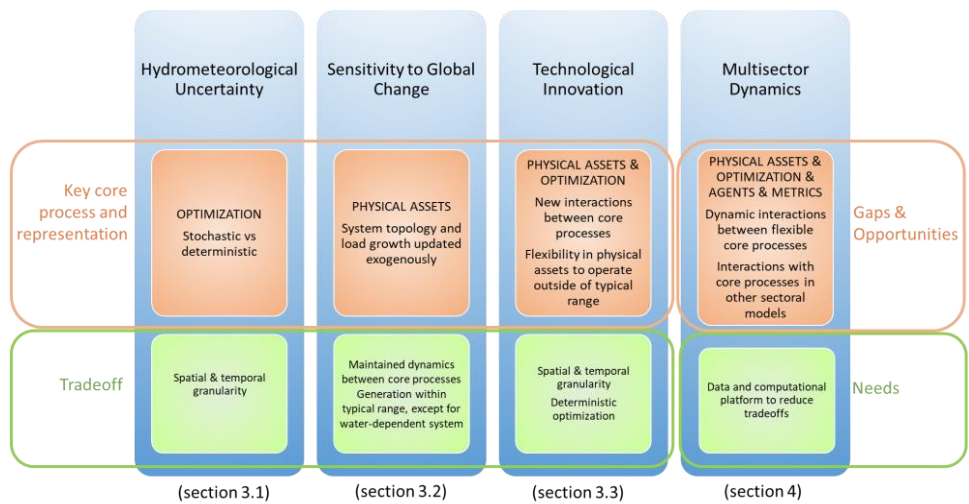

Figure 3: Summary of key modeling representations and tradeoffs per science question

## 3.1 Impacts of hydrometeorological variability.

Hydrometeorological variability (e.g., interannual, seasonal, weekly, and hourly fluctuations in weather and hydrology), including infrequent extreme events, is known to affect the operations of bulk power systems (59-64). For example, drought compromises hydropower operations (65-67) and can affect thermal power plants operations that depend on cooling water (11, 68). VRE production is sensitive to fluctuations in wind speeds and solar irradiance (59, 60), representing a growing concern as power systems increase their reliance on these resources (69, 70). Air temperatures (which influence heating and cooling demands) are the primary driver of day-to-day and seasonal changes in electricity demand (64).

By influencing both supply and demand for electricity, hydrometeorological processes can directly affect GHG emissions (60, 65, 71), wholesale electricity prices (60, 72, 73), and costs for power utilities and consumers (61, 72, 74, 75)). Deeper understanding of the effects of hydrometeorological variability on power system operations can help inform operational decision-making and feed into long-term planning efforts in the power sector and other areas. The effects of hydrometeorological uncertainty on power system operations have been explored either by subjecting a deterministic power system model to an ensemble of exogenous weather-based stressors (34, 61, 76) or by embedding stochastic representations of generation and load within the optimization itself (77, 78) (Figure 3). The tradeoffs presented by each of these two approaches are as follows:

1. Deterministic – The representation of hydrometeorological uncertainty with a deterministic optimization process typically requires a coupling with a stochastic representation of hydrometeorological processes, with either process-based or statistical models to translate weather and streamflow data into relevant power system inputs. Fifteen of the models considered, though deterministic, can support such uncertain parameters. For example, during periods of low hydropower production and/or low VRE is likely, the UC/ED algorithm operates according to the same, static decision rules and constraints that represent system operations during other periods.

2. Stochastic – Eight of the models considered directly embed hydrometeorological uncertainty into the optimization process to improve grid operational decisions. These can be further broken down by the horizon of the optimization and the objective function:

   a. Short-term decision support for grid stability and reliability: three of the eight models use two-stage dispatch optimization to mitigate generation mismatches (forecast error) between day-ahead and real-time markets, with a focus on short-term (within week) uncertainties related to weather-dependent load and renewable generating units. To reduce the computational time, all three models use a zonal network representation, a DC power flow model to approximate power flow constraints and simulate hydropower resources using a default monthly representation (i.e., fixed, HTC, PLF).

   b. Medium- to long-term decision support: The remaining five of eight models are designed to optimally schedule hydropower production. More specifically, three of five models minimize the total operating cost of a bulk power system considering both hydro and thermal generation resources under hydrometeorological uncertainties; the other two models maximize the profits of a hydro and thermal generation portfolio, considering not only hydrometeorological uncertainties but also uncertainties related to market prices for electricity. In these models, the optimization problem is usually solved for longer scheduling horizons (month, year, multiple years) under hydrometeorological uncertainty (79). To reduce computational time, all five models adopt a zonal representation, while a pipe flow or DC power flow approximation model is used to represent the power flow in transmission lines. To deal with uncertain hydrometeorological and price parameters, the models use applications of SDP and SDDP. These techniques first calculate the optimal schedules for hydropower resources in the form of expected marginal water values for each plant, given the water streamflow uncertainties in different time steps (e.g., weeks). The optimal operational decisions for the hydro and thermal resources are then made based on the marginal cost values of water (80).

The key core process that stands out to address hydrometeorological variability is optimization (Figure 3). The deterministic optimization is a passive approach to representing decision-making, and it likely underrepresents the ability of system operators to avoid adverse outcomes. On the other hand, it tends to be less computationally burdensome than the stochastic optimization, and thus facilitates analysis of system performance over a wider range of possible hydrometeorological states, including rare yet plausible events outside recorded observations.

## *3.2 Sensitivity to long-term global changes*

Key drivers of long-term (trends) global changes that subsequently affect power systems operation include but are not limited to (1) population growth and migration; (2) evolution of irrigation and transportation sectors; (3) evolution of gas infrastructure and markets; (4) climate change; and (5) extreme weather events. The effect of global change in other sectors typically influences power system models through the change in infrastructure and the changes in load and generation resources via exogenous input data sets. Global change can also propagate through endogenous processes – but as we note below, this dynamic is presently limited to hydropower.

For consistency with capacity expansion models, most operational power system models use cost optimization to evaluate the impacts of global change on the operations (17, 81-83), although it might not be the most suitable option for consistency because key drivers and agents for infrastructure expansion might not be scalable for operations (84). Population, agriculture, transportation, climate change, and extreme weather events tend to be addressed individually or in combination and are associated with changes in load profiles, magnitudes, and spatial distributions of demand (12, 85-92). Those impacts are typically represented as exogenous processes, i.e. updated time series of load and load profiles inform operational models. We noted three open-source operational models that can endogenously model other energy systems such as natural gas, heating, and transport, and coordinate their operation with power systems operations to supply the corresponding loads (e.g., electricity, heat, and gas) at minimum cost (93). The representations of the other systems remained conceptually simple where the link with the power sector is usually modeled using input-output energy conversion efficiencies and associated conversation losses. This black-box approach reduces the level of complexity to represent a system (natural gas, heating, and transport) and the number of parameters necessary to describe multisector processes.

Impacts of climate change and extreme weather events on thermo-electric power plants (i.e., capacity deratings due to cooling water shortages) have typically been represented exogenously as well, providing power systems with updated parameterization of thermo-electric capacity (11, 37, 94-96).

Climate change also alters the timing and amounts of hydropower available due to shifts in seasonal streamflow dynamics and more frequent/severe droughts and floods, competing water demands, and changes in water quality (97). Understanding the effects of long-term environmental change on hydropower operations is highly intertwined with multi-objective management of river systems for water supply, ecosystem services, and flood control. Below, we review how the first three categories of hydropower representations (all but fixed schedule) in our taxonomy can capture the impact of climate change on power systems through hydropower. While there are other ways to represent the impact of climate change on individual hydropower assets (45, 98-104)), we limit our discussion to the representations noted in our taxonomy, because they are aligned with the computational capabilities, data availability, and system needs of the power system models.

### 3.2.1 Unit commitment based on monthly hydropower potential and limited water constraints (default)

Of the 23 reviewed operational models, nine use a monthly hydropower generation potential along with HTC or PLF and minimum/maximum generation capacity inputs to guide the optimization of the hourly schedules. This representation can capture seasonal changes in water volumes but does not capture potential changes in water management. However, it allows decision agents (e.g., ISOs/RTOs, reliability coordinators) to run deterministic simulations for long scheduling horizons (typically 1 year with hourly resolution) using a detailed nodal representation of the power network (thousands of nodes) and a DC power flow approximation to inform reliability studies.

### *3.2.2. Hybrid water model and power unit commitment model (intermediate)*

Three of the 23 models use an intermediate input with an HTC and/or PLF representation and weekly to daily energy potential for representing hydropower operations. This representation provides a more realistic representation of global changes (i.e., streamflow dynamics) over river systems where river operations cannot or choose not to maintain the same level of river services. However, the inclusion of finer temporal resolutions adds to the complexity of obtaining computationally tractable solutions. This complexity is counterbalanced by using less detailed representations for modeling power network (i.e., zonal) and transmission line constraints (i.e., pipe flow).

### *3.2.3. Water model exogeneous unit commitment*

Of the 23 operational models, six are coupled with a river-routing reservoir model for modeling hydropower operations. In these models, the scheduling decisions about hydropower generation are optimized by the river-routing reservoir model instead of the operational model. This representation can capture changes in hydrologic regimes and water management rules and respond to changes in the power grid. To limit computational burden, however, the models use a zonal network representation within a confined geographic region (e.g., power utility footprint) that consists of a limited number of individual hydropower plants. The reduced domain influences the optimization process.

Of these six models, four include a dual-optimization scheme. Due to the tremendous modeling complexity of this integrated approach, several modeling assumptions are considered, including a zonal network representation (in some cases the choice of single node connection is also available), a pipe flow model for modeling transmission lines, and a weekly time resolution. In addition, in these models, appropriate formulation methods are used (typically SDP or SDDP) to overcome, to some extent, time-consumption issues; however, as the scale of the water system increases, the execution time may be prohibitive for practical applications.

Overall, as summarized in Figure 3, global change research in power system operational models stands out as limited to impact assessment studies because the fidelity and tractability of core process and the interactions between core processes are is maintained despite a change in the number of physical assets. We noted one exception when hydropower is represented with an explicit water management and river-routing process which allows to explore evolving dynamics between power operations and water management and water use competition. This maintained fidelity in the representation of core processes and their interactions is the main distinction in computational tradeoff with the next section.

## 3.3 *Evaluating adoption of new technologies*

Reducing the dependence of power systems on fossil fuels entails the diversification of technologies that compose the electric grid, including but not limited to rising shares of distributed energy resources (DERs), including flexible loads, distributed generation and energy storage resources. Some of these technologies strengthen the coupling of the power system with other resource sectors such as water resources, transportation, and agriculture, creating a need for core processes that can modify their operation in order to address changing underlying conditions.

A key component of this science question is evaluating power systems' ability to respond to relatively short-term operational and economic variabilities and uncertainties that are likely to stress the system or affect costs (47), i.e., operational flexibility, when either constrained or enhanced by stronger coupling with non-power sectors. The approach to addressing this science question contrasts with our observation of approaches to addressing global change. Operational models typically require modification of the core processes to accommodate the new technologies that can also be linked to non-power sectors. This modification is mainly related to the ability of physical assets to deviate from their existing operational state (generation, load, or transmission features).

Operational flexibility in power systems has traditionally been represented by dispatching load and generation assets whose sole purpose is to serve the power system, but operational flexibility can also be obtained by coordinating the operation of multisector systems. Relatively few studies have investigated the impacts of traditionally non-power sector technology adoption on grid operations and the need for flexibility. For example, water irrigation systems can shift the electric consumption of the water pumps to periods of low energy prices and/or to periods of high solar power generation (18, 105-107). Similarly, EVs can manage their charging strategies to align with periods that are most efficient for the power grid from an economic and reliability standpoint (108). In these studies, the operational flexibility was measured by developing appropriate flexibility metrics (108) or by measuring the energy consumption savings of the co-optimized and non-optimized cases (18).

We summarize below (and more succinctly in Figure 3) how operational flexibility is represented in core processes, specifically with respect to generation and load resources (i.e., their ability to rapidly dispatch their capacity), and optimization processes (e.g. the spatial and temporal resolution to accommodate the flexibility in physical assets).

### 3.3.1  *Generation and load resources*

Operational flexibility can be assessed in terms of how fast generation and load resources dispatch their capacity to the grid. Hence, one of the main factors toward assessing flexibility is the ability of operational models to accurately model the dynamic operation of generation and load resources.

1. Thermal generation resources: All 23 operational models can model thermal generation resources, where the dispatch decisions are based on each unit's cost function and technical characteristics. However, six of the 20 models (mainly commercial models) can also impose flexibility penalties on units with slow ramping rates and long minimum up/down times so that more flexible units are dispatched first.

2. Hydropower resources: Twelve of the 23 models can flexibly dispatch hydropower generation potential using the HTC and PLF dispatch methods to respectively respond to market signals and net load variations. The remaining 11 models represent hydropower resources using fixed generation patterns (obtained either by BAs or water-routing management models), that reflect operational constraints on the hydropower system in the hydrologic year from which the generation data were obtained. This latter representation removes any ability of hydropower plants to respond to net load or market variations associated with the integration of new technologies, unless the new flexibility needs can be quantified and the water models are rerun with this information.

3. Energy storage (ES) resources: ES resources can quickly coordinate their charge and discharge scheduling decisions with power systems' electricity prices and load variations. Hence, most operational models in our catalog (19 of 23) endogenously model the dynamic operation of ES resources by either following the load (i.e., peak shaving representation) or electricity prices (i.e., price driven representation) of the system. While ES resources are often solely constrained by power system considerations, their presence can better enable integration of technologies from other sectors such as the transportation sector (e.g., EVs). However, approaching higher levels of new technologies integration would also require seasonal storage capacities of weeks or months. In such cases, the charge and discharge decisions of some ES would need to be managed over horizons larger than operational models' typical times steps (109).

4. Distributed energy resources (DERs): DERs may include distributed generation resources (solar and wind), and distributed flexible loads (i.e., EVs, interruptible loads). Thirteen of the 23 models represent DGs resources, which, as in the large-scale VRE resources (19 of 23 models), are treated as non-dispatchable resources by operational models and as must-take, creating the need for operational flexibility. Finally, 11 operational models can rapidly dispatch their available load capacity by either reducing their electricity consumption (interruptible loads) at certain time periods or deferring the electricity demand at future times (EVs) to respond to electricity demand and price changes.

### *3.3.2 Spatial and temporal resolution*

The spatial and temporal resolutions for optimization problems set the representation of flexibility processes. In this context, a finer temporal resolution (e.g., 5 min, 15 min) can capture more of the ramping dynamics and flexibility of the available resources for grid operations, particularly because VRE and DG penetration levels increase. Overall, 11 of the 23 operational models in our catalog run sub-hourly simulations (5 min, 15 min, 30 min), ten minimize the operation cost of the system, and one maximizes the profit of a hydro and thermal generation portfolio. Because sub-hourly simulations are computationally intensive, a zonal representation and using a simplified representation for modeling power flow constraints in transmission lines (i.e., DC approximation, pipe flow model) are preferred.

## 4. Discussion – Gaps and Future Directions

Future studies evaluating new technologies and global change need to shift away from traditional paradigms, in which each sector is operated and managed in isolation and apply a more integrated operational approach. Modeling the interdependencies and co-evolution of multisector systems is expected to expose new system risks and opportunities. We posit that each core process has the potential to provide a linkage with a different sector of activity (transportation, water, etc.) which also most often have similar core processes (agents, system objective, physical assets, metrics) of their own. The taxonomy that we developed in Section 2 is expected to support model development or coupling approaches with other sectoral models. The review of computational tradeoffs in operational models in Section 3 provides critical lessons learned to be leveraged when we think about how to further develop operational models, or their coupling with other sectoral models. Computational complexity typically becomes prohibitive with greater model spatiotemporal resolution. Several tradeoffs across data availability, model fidelity, and computation burden were chosen to address the impact of uncertainty in hydrometeorological conditions, sensitivity to global changes, and adoption of new technologies into the power system (Figure 3). In this section, we identify areas in which core process representations and interactions can and must improve to enhance our understanding of how complex systems co-evolve and to develop robust opportunities for resilient systems. We discuss model fidelity in all four core processes and desired interaction and flexibility, whether those are endogenous to the operational power system model or are part of an integrated modeling platform. We then discuss the data and computational platform needs to limit or control the computation tradeoffs.

## 4.1 *Physical assets fidelity needs for analyzing multisector dynamics*

To advance the science of multisector dynamics research relying on operational models, core process fidelity needs to be addressed by future research, as follows:

1. *Fidelity of topology associated with multisectoral governance:* Bulk electricity systems operate across multiple domains including balancing regions, market regions, and the overall power grid. Irrigation and water distribution systems are defined by large river basins and watersheds for overall water management and multi-objective optimizations, and by cities for urban infrastructure. Transportation systems are typically managed by states/provinces and countries, similarly for agriculture and other human systems. While it remains unclear how much fidelity is needed to represent other sectors, the topology of the power system model should be informed by the spatial resolution of decision-making in other sectors (110).

2. *Tractability in temporal resolution*: The core processes in operational models have temporal scale of minutes or hours, up to a couple months when managing physical assets. Dominant water system dynamics occurring over the power grid spatial domain, however, have temporal scales spanning from hours to multiple years. Existing operational models fail to adjust time scales across sectors or impose time-based modeling assumptions to run coherent and observable interdependent systems. This could result in inaccurate matching of service interactions with respect to time and system state, such as matching power consumption (in Watts) to water consumption (in Liters). Temporal scales of multisector dynamics need to be further evaluated.

3. *Fidelity in representation of physical assets outside the power sector*: Multisector systems are suitable candidates for providing operational flexibility by coordinating key operational processes that consume electricity in other sectors (e.g., the timing and amounts of hydrogen produced, amount of water treated/desalinated/distributed, etc.). Such coordination could reveal significant operational benefits for interconnected infrastructures like a reduction in operating costs and the provision of flexibility services. However, existing operational models do not support an integrated operational framework that can explicitly model each sector's operational constraints. The flexibility that a sector can exchange with another sector is currently limited by the physical constraints of both sectors. For example, the flexibility of the transportation network is subjected to decisions associated with EV drivers' behaviors (delays, deadlines, stops, etc.), traffic patterns (i.e., traffic congestion), and the specific spatial topology of the transportation network (e.g., number and type of chargers, etc.). A failure to capture multisector constraints could lead to an over or underestimation of the available flexibility or even to decisions that might adversely affect the operational objectives of the coupled infrastructures.

## 4.2 New system objectives and metrics associated with multisector dynamics

In addition to improved model fidelity in the representation of interactions between the power grid assets and other sectors', it is critically important to have robust assessment methodologies and indicators that can evaluate multisector dynamics. In particular, optimization and metrics that allow for interpretable and generalizable results across spatial domains, and metrics that allow for characterizations of time dynamic interactions and vulnerabilities across sectors, are needed. Within this context, additional research is needed to develop metrics that can inform multisector entities (e.g., water utility operators, gas operators, hydropower operators) about how they can hypothesize the impact of energy consumption, costs, and revenues on their operations. These optimization and associated metrics should be able to capture the level of interconnectedness between sectors to reflect performance relative to different temporal (e.g., seasonal) and spatial (state, regional, national level) levels:

1. *Consistency between multi-objective functions*: Existing operational models do not consider the potential for multiple competing objectives across grid participants and sectors. Examples include hydropower plant electricity producers that aim to maximize profits versus ISOs/RTOs that aim to minimize operating costs or additionally reconciling these monetary objectives with non-monetary goals. Other examples include competing water objectives in multipurpose reservoir operations (hydropower generation vs irrigation) or between multisector infrastructures and associated infrastructure operators. In non-electricity sectors, the infrastructure's operation is not only based on the lowest cost practices but also on ensuring reliable provision of services. More comprehensively analyzing global change and technology innovation to explore tradeoffs should include valuations of non-energy commodities (water, transportation, gas) to better represent the value of multisector services. Consistent optimization leads to new metrics as discussed below.

2. *Environmental impacts from linked, non-power infrastructure:* In addition to the operations of infrastructure in non-power sectors contributing to changes in environmental impacts associated with the power sector, the reverse can also occur: changes in power sector operation can affect the environmental impacts associated with linked infrastructure in non-power sectors. For example, the response of hydropower facilities to better balance the dynamics of wind and solar may result in changes in downstream water quality due to sudden water releases changing the shape of the downstream hydrograph (111). Performance metrics that account for effects such as these alongside power sector metrics will provide a more complete picture of the environmental impacts associated with multisector dynamics and enable planners to avoid unintended consequences.

3. *Reliability impacts in non-power systems:* Uncertainties associated with other external forcing (e.g., climate change) and model representation choices can propagate to non-power sectors. For example, a contingency on the electric grid due to extreme weather or climate change can result in unmet loads and blackouts, which can subsequently affect the physical reliability of equipment in other sectors (e.g., water treatment plants) to operate at their design conditions. Metrics that capture how risk and uncertainty propagate across the operations of multiple sectors are critical for characterizing and alleviating undesirable multisector dynamics.

Considering the wide range of core processes defined in our taxonomy, different types of potential metrics can refer to different science questions. For example, if a process-based approach is used to enhance the fidelity of representing other sectors and their interactions, we recommend developing fidelity metrics. The science questions we reviewed would then be complemented with metrics that track how the different sectors co-evolve. If statistical and machine learning approaches are used to represent the connection with other systems, core process-specific fidelity models might not be adequate, but the multi-system objectives would still need to be evaluated. Our taxonomy is relatively generalizable to other human systems such as transportation, water management, natural gas, and economics.

## 4.3 Multisectoral markets and agents

In addition to the need for developing multisector optimization and metrics, we identified the need for exploring new market products and mechanisms as key elements for facilitating collaboration among multisector stakeholders and integrating non-power sector modeling objectives. Such efforts could include:

1. Facilitate demand-side participation of multisector systems in electricity markets, including the development of effective price signals.

2. Develop new market products for services provided by multisector systems, such as seasonal flexibility (driven by environmental changes such as water availability) and demand-side electrification (driven by urbanization and migration trends).
3. Develop market valuation methods for commodities other than electricity such as water, gas, and food supplies.
4. Expand beyond the traditional ancillary and energy services to remunerate a comprehensive suite of services, including markets for emergency response services during resilience restoration strategies.

In response to new market products, new business models need to be developed to access the tariff structures that maximize flexibility over short- and long-time frames. These models could lead to a paradigm that causes a radical change in the power market architecture as we know it today, replacing traditional downstream marketing of power by increasing reliance on multisector interactions on the demand and supply side of the equation. In this direction, recent research focus is given on developing models for transactive energy (TE) markets in power distribution systems that supply the demand of other sectors. Under the TE paradigm, distribution system operators (energy suppliers) and the sectors they serve (load consumers) could barter over the proper way to solve their mutual problems, and settle on the proper price for their services, in close to real time (112). Such settlement procedures require appropriate renumeration algorithms that leverage statistical and machine learning methodologies and optimize reward and market structures (113).

## 4.4 Computational and data exchange requirements

The broader integration of multisector processes in power system operations would also require exchanging a plethora of market (e.g., electricity prices and tariffs), environmental (e.g., wind, solar, water), and system state (e.g., voltages, pressurized) information between multisector systems. To achieve such integration, it is crucial to establish appropriate communication protocols that can ensure the reliable and safe information and data exchange between multisector operators, while handling large amounts of data and translating them into understandable decision actions (e.g., opportunities to provide demand flexibility). Therefore, the integration of intelligent metering, telecommunication, control, and automation systems for each sector and associated controllable equipment (e.g., EV charging stations, water treatment facilities, pump stations, etc.) is critical to building "intelligent" infrastructures.

Finally, while this discussion remains open to approaches taken to evaluate adequate sectoral model fidelity for different science questions and the associated computational burden, an even more immediate challenge is the lack of high-quality data pertaining to each sector that span spatial (e.g., the geographic coverage of the coupled infrastructure) and temporal (e.g., irrigation needs for each crop, etc.) dynamics.

## 5. Summary and Conclusion

We presented a novel taxonomy of power system operations models, with a focus on core processes. The modeling capabilities and features of 23 operational models were compartmentalized into four core process representations: (1) model objectives and purpose; (2) physical grid assets (transmission, generation resources, loads, and energy storage resources); (3) institutions and decision agents; and (4) performance metrics. The process-driven taxonomy was leveraged to report on the critical core processes and associated computational tradeoffs observed in existing studies addressing three common science questions around hydrometeorological variability, global change, and the adoption of new technologies. The computational tradeoffs review provided a basis for realistic recommendations on model development needs and research to advance our understanding of complex interactions with other sectoral activities. This multisector dynamic research direction will help represent multisector dependencies which expose the risks and opportunities in power system operations.

More specifically, we identified challenges in computation times, data availability, performance metrics, and incorporating multisector dynamics in operational models. A clear challenge for operational models is the proper consideration of the interdependencies and co-evolution of multisector systems, particularly related to the fidelity in representing multisector constraints, multi-objective representations, and their spatial and temporal resolutions, which do not appear to be considered fully in most operational models. Our analytics allowed to identify the different core processes that require a coordinated evolution in fidelity and support the development of multisector systems, whereby each system optimally interacts with each other at various levels.

Finally, we recommend for future literature using operational models to be more transparent about the core process representations, with the aim to clarify how generalizable the approach and models are to other regions or science questions. The region and science question-specific representations of core processes and their combinations reflect not only the maturity of the science but also data availability and existing tradeoffs about model fidelity and computational requirements. This blueprint provides guidance for introducing the core processes of the model and support its representation for addressing the science question and for discussing findings in a reproducible way.

## Acknowledgements

This research was supported by the U.S. Department of Energy, Office of Science, as part of research in the MultiSector Dynamics, Earth and Environmental System Modeling Program. This work was authored by the Pacific Northwest National Laboratory, managed by Battelle (contract no. DE-AC05-76RL01830) for the U.S. Department of Energy.